\documentclass[aps,prd,amsmath,amssymb,twocolumn,preprintnumbers]{revtex4}
\pdfoutput=1
\usepackage{graphicx}
\usepackage{color}
\newcommand{\beq}{\begin{eqnarray}}
\newcommand{\eeq}{\end{eqnarray}}

\newcommand{\bmp}{\noindent\begin{minipage}{16cm}}
\newcommand{\emp}{\end{minipage}\vskip 7mm} 

\usepackage{dcolumn}
\usepackage{bm}
\usepackage{bbm}
\usepackage{subfigure}
\usepackage{pxfonts}

\usepackage{ulem}

\definecolor{rossoCP3}{cmyk}{0,.88,.77,.40}

\def\lsim{\mathrel{\rlap{\lower4pt\hbox{\hskip1pt$\sim$}}
    \raise1pt\hbox{$<$}}}                
\def\gsim{\mathrel{\rlap{\lower4pt\hbox{\hskip1pt$\sim$}}
    \raise1pt\hbox{$>$}}}                

\newcommand{\be}{\begin{eqnarray}}
\newcommand{\ee}{\end{eqnarray}}

\preprint{CP3-ORIGINS-2010-50}

\usepackage{fancyhdr}
\begin{document}
\title{\Large  \color{rossoCP3}   Beta Function and  Anomalous Dimensions}
\author{Claudio {\sc Pica}$^{\color{rossoCP3}{\varheartsuit}}$}
\email{pica@cp3.sdu.dk} 
\author{Francesco {\sc Sannino}$^{\color{rossoCP3}{\varheartsuit}}$}
\email{sannino@cp3.sdu.dk} 
\affiliation{
$^{\color{rossoCP3}{\varheartsuit}}${ CP}$^{ \bf 3}${-Origins}, 
Campusvej 55, DK-5230 Odense M, Denmark.}
\begin{abstract}
We demonstrate that it is possible to determine  the coefficients of an all-order beta function linear in the anomalous dimensions using as data the two-loop coefficients together with the first one of the anomalous dimensions which are universal. The  beta function allows to determine the anomalous dimension of the fermion masses at the infrared fixed point, and the resulting values compare well with  the lattice determinations.\\[.1cm]
{\footnotesize  \it Preprint: CP$^3$-Origins-2010-50}
\end{abstract}

\maketitle
\thispagestyle{fancy}

Understanding the non-perturbative dynamics of gauge theories of fundamental interactions constitutes a formidable challenge. Recently a considerable effort has been made to unveil the large distance conformal dynamics of  these theories, see  \cite{Sannino:2009za} for a recent review. 

The goal here is to prove the existence of an all-orders beta function similar, in shape, to the one provided by Ryttov and Sannino (RS) in \cite{Ryttov:2007cx}.

We consider a generic vector-like gauge theory with $N_r$ Dirac fermions transforming according to distinct representations $r$ of the underlying gauge group. The beta function of any gauge theory is a gauge independent quantity and therefore can only depend on gauge invariant quantities. Besides, gauge invariance must be respected at each order in perturbation theory. Since the anomalous dimensions of the fermion masses are gauge independent the beta function can depend on them. We can therefore always write:
\begin{equation}
{\beta(\alpha)} =  f(\alpha, N_r\, \gamma_r \ , \gamma_g) \ ,~~{\rm with} ~~ r = 1,\ldots, p \ ,
\end{equation}
where 
\begin{equation}
\beta(\alpha) = \frac{\partial \alpha}{\partial \ln \mu} \ , \quad {\rm and} \quad  \gamma_r = - \frac{\partial \ln m_r}{d \ln \mu} \ .
\end{equation}
 $m_r$ is  the Dirac mass of each fermion species.  $\gamma_g$  is the gluon wave function renormalization anomalous dimension satisfying the important relation:
\begin {equation}
\gamma_g = \frac{\beta}{\alpha} \ ,
\label{background}
\end{equation} 
in the background field method \cite{DeWitt:1967yk,Abbott:1981ke,Abbott:1980hw}. Note that the background field method does not fix uniquely the gauge. More precisely the quantities $\beta$, $\gamma_r$s and $\gamma_g$ do not depend on the gauge fixing parameter $\xi$ while, on the the other hand the anomalous dimensions of the wave functions of the fermion fields do depend on this parameter. It will become soon clear that these anomalous dimensions are the minimal set of gauge invariant quantities needed to determine the beta function. 

We assume  $f(\alpha, N_r\, \gamma_r \ , \gamma_g)$  to be such that  $\beta(\alpha)/\alpha^2$ is linear in the anomalous dimensions:
\begin{equation} 
\frac{\beta(\alpha)}{\alpha} = - \frac{\alpha}{2\pi} \left[ a + \sum_{r=1}^p a_r\, {N_r}\gamma_r  - a_g \gamma_g  \right] \ ,
\label{beta0}
\end{equation}
with $r= 1,\ldots, p$  labeling matter transforming according to distinct representations of the underlying gauge group.  We stress that the coefficients $a$, $a_r$ and $a_g$ are independent on $\alpha$ by assumption. We will prove below that, in perturbation theory, a scheme exists for which Eq.~\eqref{beta0} is valid.
 
Using \eqref{background} we find: 
\begin{equation}
\frac{\beta(\alpha)}{\alpha} = - \frac{\alpha}{2\pi}\frac{a + \sum_{r=1}^p a_r \, {N_r}\gamma_r   }{1 - \frac{\alpha}{2\pi} a_g} \ .
\label{betaao}\end{equation}

We will now {\it show} that  it is possible to determine the $p+2$ unknown coefficients $a$, $a_r$ and $a_g$ using the universal coefficients of the two-loop beta function together with the universal coefficient of the anomalous dimension of the mass for each representation. We henceforth introduce the two-loop beta function: 
\begin{equation}
\frac{\beta^{2L}}{\alpha} = -\frac{\alpha}{2\pi} \left[\beta_0 + \frac{\alpha}{2\pi} \beta_1 \right] \ , 
\end{equation}
with $\beta_0$ and $\beta_1$ the two universal coefficients: 
\begin{eqnarray}
\beta_0  & = &\frac{11}{3}C_2[G] - \frac{4}{3}\sum_{r=1}^pT[r] N_r \ ,  \\ 
\beta_1 & = & \frac{17}{3}C_2[G]^2 - \frac{10}{3}C_2[G]\sum_{r=1}^pT[r] N_r  - 2 \sum_{r=1}^p C_2[r] T[r]N_r \ . \nonumber \\
&& 
\end{eqnarray}
 $C_2[r]$ is the quadratic Casimir and $T[r]$ are the normalizations of the generators in the representation $r$, and $G$ indicates the adjoint representation. 
The explicit expressions for $C_2[r]$ and $T[r]$ can be found in \cite{Sannino:2009za}.
Expanding the all-orders beta function to two-loop and from the matching of the coefficients we find:
\begin{eqnarray}
a = \beta_0 \ , \quad 
a_g \beta_0 + \sum_{r=1}^p a_r \, k_r\, N_r  =\beta_1 \ ,
\label{system}
\end{eqnarray} 
with $k_r =3 C_2(r)$  a universal constant associated to the anomalous dimensions as follows:
\begin{equation}
\gamma_r = \frac{\alpha}{2 \pi} k_r + {\cal O}\left(\alpha^2 \right) \ .
\label{gammar}
\end{equation}
The remaining $p+1$ coefficients, $a_g$ and $a_r$ in \eqref{beta0} can be determined from \eqref{system}. We start by observing that, in general, the coefficients $a_g(N_r)$ and $a_r(N_r)$ can depend on the the number of fermion species $N_r$.
We start by considering the pure Yang-Mills case for which the only unknown coefficient $a_g$ is immediately deduced from \eqref{system} by setting $N_r=0$ to be: 
\begin{equation}
a_g(0)   = \frac{\beta_1^{YM}}{\beta_0^{YM}} = \frac{17}{11}C_2[G]   \ ,
\label{agsolution}
\end{equation}
and therefore the pure Yang-Mills beta function becomes: 
\begin{equation}
\frac{\beta_{\rm YM}(\alpha)}{\alpha} = - \frac{11}{3}\frac{\alpha}{2\pi}\, \frac{C_2[G]}{1 - \frac{\alpha}{2\pi} \frac{17}{11}C_2[G] } \ .
\label{betaaoYM}\end{equation}
This is the RS beta for the pure Yang-Mills. The resulting mass gap and the study of the poles of this theory have been analyzed recently in \cite{Sannino:2010ue}. We also understand why the running of this beta function captures  so well the lattice simulations performed using the Schr\"odinger functional scheme as shown in \cite{Ryttov:2007cx}. The reason is that the present beta function and the one derived on the lattice are both based on a background field method implying automatically the relation \eqref{background} in the two cases.

To show how to derive the other coefficients we consider first the case of a single matter representation $R$ for which the constraint in \eqref{system} reads:  
\begin{equation}
a_g \beta_0 + a_R \, k_R N_R = \beta_1. 
\end{equation}
 We therefore determine the coefficient $a_R$, at the value $\overline{N}_R$ for which asymptotic freedom is lost, i.e. $\beta_0=0$, to be: 
\begin{equation}
a_R ( \overline{N}_R ) = \frac {\beta_1(\overline{N}_R)}{k_R \overline{N}_R} \ .
\label{arsolution}
\end{equation}
The two conditions \eqref{agsolution} and \eqref{arsolution} did not have, in principle, to lead to a universal solution valid for any number of matter fields. However, it is straightforward to show that a solution to \eqref{system} is obtained if $a_g$ and $a_R$ are constant, i.e. do not depend on $N_R$, and equal to the values given by \eqref{agsolution} and \eqref{arsolution}.
In fact one can consider the two cases in which either $a_R$ does not depend on $N_R$  or the case in which $a_g$ is constant. Surprisingly these two limits lead to the same solution. The reason lies in the linearity of the two-loop beta function coefficients with the number of flavors $N_R$.
Thus the solution with $a_R$ and $a_g$ constants independent on $N_R$ is the most natural choice.
We note, however, that it is not the only logical possibility. In fact other solutions can be easily built, for example, by choosing any function $a_R(N_R)$ such that $a_R ( \overline{N}_R )=a_R ( 0 )={\beta_1(\overline{N}_R)}/{(k_R \overline{N}_R)}$, and then choosing $a_g(N_R)$ to satisfy \eqref{system}. The constraints on $a_R$ guarantees the solution to be regular at $N_R=0$ and $\overline{N}_R$ as we will discuss in more detail below.
In the following we will only consider the solution with $a_R$ and $a_g$ constant and study its consequences.

Given the two coefficients $a_g$ and $a_R$ the all-orders beta function \eqref{betaao}, for a single representation $R$, is determined to be
\begin{equation}
\frac{\beta(\alpha)}{\alpha} = - \frac{\alpha}{2\pi}\frac{\beta_0 +  \frac{\beta_1(\overline{N}_R)}{\overline{N}_R k_R} N_R\gamma_R   }{1 - \frac{\alpha}{2\pi} \frac{\beta_1^{YM}}{\beta_0^{YM}}} \ .  
\label{betaps}\end{equation}

The expression above is, in shape, identical to the RS beta function \cite{Ryttov:2007cx} and, for $N_R=0$, matches the Yang-Mills result. 
However for $N_R\to \overline{N}_R$ the RS does not reproduce the value of the anomalous dimension at the perturbative infrared fixed point. This occurs because the RS solution does not satisfy the condition \eqref{arsolution}. We will show below that \eqref{arsolution} is, in fact, a necessary condition if the perturbative expansion of the anomalous dimension at the fixed point has to be recovered.

We note that, in general, if either \eqref{agsolution} or \eqref{arsolution} are not satisfied, the coefficients $a_g$ or $a_R$ will have pole singularities at $N_R=0$ or $N_R=\overline{N}_R$.
In fact, for example, it is easily seen from \eqref{system} that if \eqref{arsolution} is not satisfied then $a_g(N_R)\sim 1/\beta_0$ for $N_R\to \overline{N}_R$. 
 This generalizes to any number of matter representations.

Note also that for the solution proposed here the denominator does not depend on the number of flavors. Finally the value of the anomalous dimension at the fixed point is corrected with respect to the RS result. We rewrite the beta function after having evaluated the different coefficients:
\begin{equation}
\frac{\beta(\alpha)}{\alpha} = - \frac{\alpha}{6\pi} \frac{11 C_2[G] - 2 T[R] N_R(2 +  \Delta_R \gamma_R)  }{1 - \frac{\alpha}{2\pi} \frac{17}{11}C_2[G]} \ ,
\end{equation}
with 
\begin{equation}
\Delta_R = 1 + \frac{7}{11} \frac{C_2[G]}{C_2[R]} \ .
\label{eq:deltar}
\end{equation}
Interestingly in \cite{Antipin:2009dz} the same form of the beta function was also suggested among a one-parameter family of solutions of \eqref{system}, partially motivated from holography and assuming the RS shape of the beta function. It is worth stressing that our main point here is to {understand the assumptions behind}   a field theoretical justification for the RS form of the beta function and the uniqueness of the solution introduced here.

The new analytical expression of the anomalous dimension of the mass at the infrared stable fixed point is obtained by setting the beta-function to zero and reads:
\begin{equation}
\gamma_R=-\frac{\beta_0(N_R)}{\beta_1(\overline{N}_R)}\frac{\overline{N}_R}{N_R}k_R =\frac{11 C_2[G] - 4 T[R] N_f}{2 N_R T[R]  \left( 1 + \frac{7}{11} \frac{C_2[G]}{C_2[R]} \right)} \ .
\label{eq:gammar}
 \end{equation}
The RS result is instead:  
\begin{equation}
\gamma^{\rm RS}_R =\frac{11 C_2[G] - 4 T[R] N_R}{2 N_R T[R]} \ .
 \end{equation}
It is useful to compare the two anomalous dimensions:  
\begin{equation}
\gamma_R = \frac{11}{11+ 7  \frac{C_2[G]}{C_2[R]} }\gamma^{\rm RS}_R \ .
\label{gammaRSw}
\end{equation}
This shows that the corrected anomalous dimension at the fixed point is smaller than the one predicted earlier, which agreed roughly with the Schwinger-Dyson results. 

We now show that if $a_r$ differs from \eqref{arsolution} the anomalous dimension {\it at the fixed point}  obtained via \eqref{beta0} does not  reproduce the perturbative value. The latter is obtained using the two-loop beta function and the one-loop gamma function which gives:
\begin{eqnarray}
\gamma_R^{\rm perturbative}=-\frac{k_R}{\beta_1(\overline{N}_R)}\beta_0 + \mathcal{O}(\beta_0^2)\ .
\label{eq:gammaper}
\end{eqnarray}
 On the other hand from the proposed beta function we obtain \eqref{eq:gammar}:
\begin{eqnarray}
\gamma_R=-\left(a_r(\overline{N}_R)\overline{N}_R\right)^{-1}\beta_0 + \mathcal{O}(\beta_0^2)\ .
\label{eq:gammaao}
\end{eqnarray}
Requiring, for consistency, equations \eqref{eq:gammaper} and \eqref{eq:gammaao} to agree in perturbation theory allows to recover exactly \eqref{arsolution}. This new condition implies that only the regular solutions, i.e. non singular in $\beta_0$, are acceptable when solving \eqref{system}. 

Next, we would like to show that it is always possible to find a renormalization scheme in which \eqref{betaps} holds. To this goal we introduce the general perturbative expansion for the anomalous dimensions and the beta function: 
\begin{eqnarray}
\frac{\beta(\alpha)}{\alpha} & = & - \frac{\alpha}{2\pi} \sum_{n=0}^{\infty} \left(\frac{\alpha}{2\pi}\right)^n \beta_n \ , \\
\gamma_R(\alpha) & = &  \sum_{n=1}^{\infty} \left(\frac{\alpha}{2\pi}\right)^n \gamma_{n-1}  \ ,
\end{eqnarray}
and by inserting the equations above in \eqref{betaps}  we derive the infinite set of relations: 
\begin{equation}
\gamma_ {n-1} = \frac{\beta_n - \beta_{n-1}\, a_g }{N_R a_R} \ , \quad  n\geq 1  \ , 
\label{relation}
\end{equation}
with $\gamma_0 = k_R = 3 C_2(R)$. 
The transformation laws between the beta function and the anomalous dimensions above and the ones in another mass-independent renormalization scheme  
  $\tilde{\beta}(\tilde{\alpha})$ and $\tilde{\gamma}_R(\tilde{\alpha})$ are:
\begin{eqnarray}
\tilde{\alpha} & = & \sum_{n=1}^\infty h_n {\alpha}^n \ , \quad {\rm with} \quad h_1 = 1 \ ,  \\ 
\tilde{m}  & =  & m\, z_{m} \left(\alpha \right)=m   \sum_{n=0}^\infty \ell_n {\alpha}^n \ , ~ {\rm with} ~ \ell_0 = 1 \ ,  \\ 
\beta(\alpha) & = & \frac{\partial {\alpha}}{\partial{\tilde{\alpha}}}     \tilde{\beta}(\tilde {\alpha})\ , \label{betaym} \\
 \gamma_R(\alpha)& = &\tilde{\gamma}_R(\tilde {\alpha})  -   \tilde{\beta}(\tilde {\alpha}) \frac{\partial {\ln z_m}}{\partial{\tilde{\alpha}}} \ .
\label{schemechange}
\end{eqnarray} 
$\gamma_0$, $\beta_0$ and $\beta_1$ are unchanged  being universal. The other coefficients do depend on the transformation and we report the explicit form of the first coefficients:  
\begin{eqnarray}
\beta_2  & = &  4\pi^2 \beta_0 ( h_2^2 - h_3) + 2\pi h_2  \beta_1 + \tilde\beta_2  \ , \\
\gamma_1 & = & 2\pi \ell_1  \beta_0 + 2\pi h_2  \gamma_0 + \tilde\gamma_1 \ .
\end{eqnarray}
Imposing  the  relation \eqref{relation} between $\gamma_1$, $\beta_1$ and $\beta_2$ yields: 
\begin{eqnarray} 
 4\pi^2 \beta_0 (h_2^2 -h_3)  - 2\pi \, a_R N_R \ell_1 \beta_0 - a_g \beta_1 + 2\pi h_2 \beta_1 +  \nonumber \\ 
 + \tilde{\beta}_2 - 2 \pi a_R h_2 N_R \gamma_0 - a_R N_R \tilde{\gamma}_1 =0 \ .
\end{eqnarray}
 To each successive order in perturbation theory, two more coefficients (one from $\alpha$ and one from $z_m$) appear and only one relation  coming from \eqref{relation} should be imposed. Therefore we can always write \eqref{betaps} as a series in perturbation theory. 

The new anomalous dimensions at the fixed point in \eqref{eq:gammar} are in better agreement with the lattice determinations, which tend to be smaller than the RS estimate.
 {The value of the anomalous dimensions at the fixed point is a physical quantity and therefore does not depend on the renormalization scheme. This fact can be directly verified using Eq.~\eqref{schemechange}.}
We quote below the value obtained for different theories on the lattice which have to be taken \textit{cum granum salis} given that they are still subject to large systematic errors difficult to estimate. 

 The anomalous dimension of the mass for the Minimal Walking Technicolor theory \cite{Sannino:2004qp,Dietrich:2006cm} corresponding to an $SU(2)$ gauge theory with two Dirac flavors in the adjoint representation is now predicted to be $\gamma_{\rm MWT} = 11/24 \simeq 0.458$ rather than $0.75$ as predicted earlier. 
 This result compares well with the latest lattice result $\gamma_{\rm MWT}^{\rm lattice} = 0.49(13)$ \cite{Catterall:2010du}, while a more conservative lattice estimate indicates the fixed value to be smaller than $0.56$ \cite{Bursa:2009we}. The anomalous dimension of the mass for the Next to Minimal Walking Technicolor theory (NMWT) corresponding to an $SU(3)$ gauge theory with fermions in the two index symmetric representation is predicted to be $\gamma_{\rm NMWT} =143/173\simeq 0.827$ which is closer to the one obtained via first principle lattice simulations in \cite{DeGrand:2010na}. 
 
 For $3$ colors and $10$ flavors in the fundamental representation we find  $\gamma_{\rm Fund} \simeq 0.53$ versus $1.3$ obtained via the RS result.

\begin{figure}[t]
\includegraphics[width=\columnwidth]{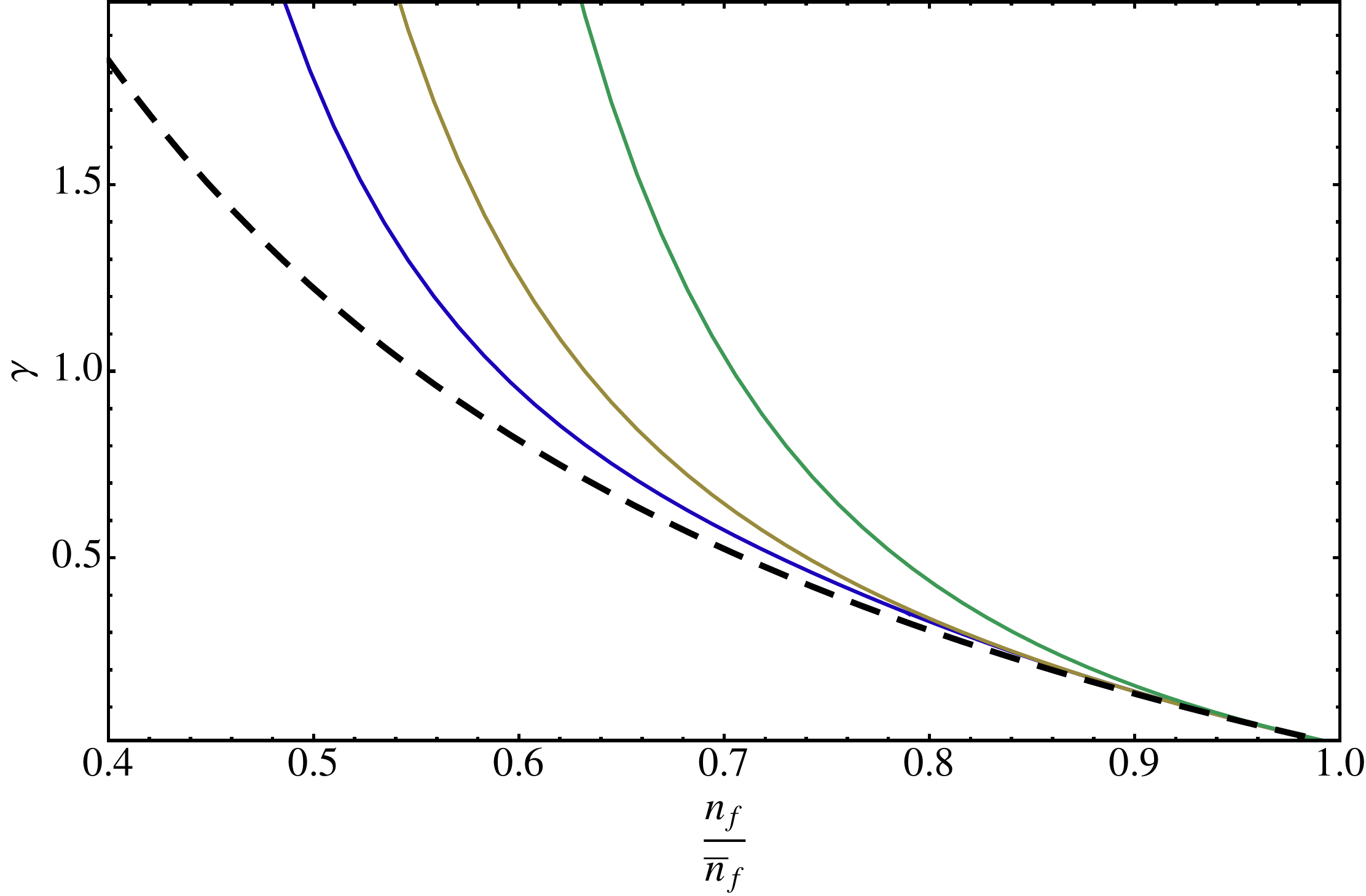}
\caption{Comparison of the predicted value of anomalous dimension at the fixed point from the proposed all-order $\beta$ function and the one coming from perturbation theory up to 4 loops.  As a representative gauge theory we used here the $SU(2)$ gauge group with fermions in the adjoint representation. Similar curves are obtained for other gauge theories. From the top curve down: 2, 3 and 4-loop perturbative results. Dashed line: anomalous dimension coming from the all-order $\beta$ function.\label{fig.gamma}}
\end{figure}

 {To provide further support to our results we compare the prediction for the anomalous dimension at the fixed point given in Eq.~\eqref{eq:gammar} with the results of perturbation theory, known up to 4 loops. We find a remarkable agreement between the two results. This is a surprising agreement given that our proposal makes use only of the 2-loop universal coefficients of the $\beta$ and $\gamma$ functions. To better appreciate the striking agreement with perturbation theory, we plot in Fig.~\ref{fig.gamma} the fixed point value of $\gamma_R$ from Eq.~\eqref{eq:gammar} and from perturbation theory at 2, 3 and 4-loop, in the case of $SU(2)$ gauge theory with adjoint fermions. The prediction coming from the proposed $\beta$-function nicely envelops the perturbative results.  Similar results are obtained for other gauge theories. We tabulate in the Appendix the values of $\gamma_R$ corresponding to the gauge groups $SU(2)$, $SU(3)$ and $SU(4)$ with fermions in the fundamental and the two indices representations.}

Using the linearity in the number of flavors of the first two-coefficients of the beta function it is straightforward to generalize the beta function to any  number $p$ of fermionic species, the result being:
\begin{equation}
\frac{\beta(\alpha)}{\alpha} = - \frac{\alpha}{6\pi} \frac{11 C_2[G] - 2 \sum_{r=1}^p T[r] N_r(2 +  \Delta_r \gamma_r)  }{1 - \frac{\alpha}{2\pi} \frac{17}{11}C_2[G]} \ ,
\end{equation}
with $\Delta_r$ given by \eqref{eq:deltar}. 
Our results generalize immediately to the supersymmetric case \cite{Novikov:1983uc} by simply replacing the corresponding quantities with the supersymmetric ones in \eqref{agsolution} and \eqref{arsolution} and obtain, for a single representation: 
\begin{equation}
\frac{\beta^S(\alpha)}{\alpha}  = - \frac{\alpha}{2\pi} \frac{\beta_0^S + 2 T[R]N_R \gamma_R}{1 - \frac{\alpha}{2\pi} C_2[G]} \ ,
\end{equation}
\noindent
where the anomalous dimension of the mass is  defined with an overall sign difference with respect to the nonsupersymmetric case and the $S$ apex means supersymmetric, for example, $\beta_0^S = 3 C_2(G) - 2 T[R]N_R$.  We find remarkable that the beta function is shape invariant when going from the nonsupersymmetric to the supersymmetric parent theory. This was not the case for the RS beta function. 

We have argued that the beta function for any nonsupersymmetric  vector like gauge theory with fermionic matter as well as its supersymmetric version can be written in the universal form \eqref{betaps}. The latter is obtained solely in terms of the universal coefficients of the two-loop beta function and the universal coefficient of the first term in the coupling  constant expansion of the anomalous dimension of the mass operator.

From \eqref{betaps} it is possible to read the values of the anomalous dimension of the mass inside the conformal window, which is found to be in fair agreement with the numerical estimates obtained via numerical simulations.

 We thank Oleg Antipin, Marco Nardecchia, Thomas A. Ryttov and Joseph Schechter for helpful comments. 

\appendix

\section{Perturbative versus all orders results}

 {In this section we report the values of the fixed point anomalous dimension of the fermion mass for different gauge theories as obtained from Eq.~\eqref{eq:gammar} $\gamma^*_{AO}$, and from perturbation theory at 2-loop $\gamma^*_{2}$, 3-loop $\gamma^*_{3}$ and 4-loop $\gamma^*_{4}$. We also report the scheme dependent value of the coupling constant $A^* = \alpha^*/4\pi$ corresponding to the zero of the perturbative $\beta$ function at 2, 3 and 4-loop. The perturbative expressions are obtained in the $\overline{MS}$ scheme \cite{vanRitbergen:1997va,Vermaseren:1997fq,Pica:2010xq}. The gauge theories considered in this Appendix are based on gauge group $SU(2)$, $SU(3)$ or $SU(4)$ with fermions in the fundamental, adjoint, 2-index symmetric or anti-symmetric representation. }
 
It is clear that for sufficiently low number of flavors, as expected, the four-loop approximation breaks down. This is the case, for example of the 7 and 6 flavors for the $SU(2)$ with fundamental fermions where one observes wild fluctuations of the fixed point couplings and anomalous dimensions at different orders. Notably the all-orders results are, on the other hand, stable.

\begin{table}[h!]
\begin{tabular}{ccccccccc}
\hline\multicolumn{9}{c}{SU(2) Fundamental} \\ \hline
$n_f$ & $n_f/\bar{n}_f$ & $A_ 2^*$ & $\gamma _ 2^*$ & $A_ 3^*$ & $\gamma _ 3^*$ & $A_ 4^*$ & $\gamma _ 4^*$ & $\gamma _ {\text{AO}}^*$\\
6.000 & 0.545 & 0.909 & 33.200 & 0.131 & 0.925 & 0.191 & -4.020 & 0.618 \\
7.000 & 0.636 & 0.225 & 2.670 & 0.084 & 0.457 & 0.096 & 0.033 & 0.424 \\
 8.000 & 0.727 & 0.100 & 0.752 & 0.055 & 0.272 & 0.060 & 0.204 & 0.278 \\
 9.000 & 0.818 & 0.047 & 0.275 & 0.033 & 0.161 & 0.035 & 0.157 & 0.165 \\
 10.000 & 0.909 & 0.018 & 0.091 & 0.016 & 0.074 & 0.016 & 0.075 & 0.074 \\
 11.000 & 1.000 & 0.000 & 0.000 & 0.000 & 0.000 & 0.000 & 0.000 & 0.000 \\
\hline
\end{tabular}
\caption{Comparison between different determinations of anomalous dimension of the mass for the SU(2) gauge theory with $N_f$ with fundamental fermions. The anomalous dimensions $\gamma^*_2$, $\gamma^*_3$$\gamma^*_4$ are the perturbative result at 2, 3 and 4-loop respctively while $\gamma^*_{AO}$ correspond to Eq.~\eqref{eq:gammar}. For the perturbative results we also report the corresponding (scheme-dipendent) value of the zero of the $\beta$ function ($A^*=\alpha^*/4\pi$ ), at 2, 3 and 4-loop respectively and indicated with $A^*_2$, $A^*_3$ and $A^*_4$ .\label{TBL1}}
\end{table}

\begin{table}[h!]
\begin{tabular}{ccccccccc}
\hline\multicolumn{9}{c}{SU(3) Fundamental} \\ \hline
$n_f$ & $n_f/\bar{n}_f$ & $A_ 2^*$ & $\gamma _ 2^*$ & $A_ 3^*$ & $\gamma _ 3^*$ & $A_ 4^*$ & $\gamma _ 4^*$ & $\gamma _ {\text{AO}}^*$\\
 9.000 & 0.545 & 0.417 & 19.800 & 0.082 & 1.060 & 0.085 & -0.143 & 0.685 \\
 9.500 & 0.576 & 0.255 & 8.030 & 0.070 & 0.813 & 0.074 & 0.048 & 0.606 \\
 10.000 & 0.606 & 0.176 & 4.190 & 0.061 & 0.647 & 0.065 & 0.156 & 0.535 \\
 10.500 & 0.636 & 0.129 & 2.500 & 0.053 & 0.528 & 0.057 & 0.218 & 0.470 \\
 11.000 & 0.667 & 0.098 & 1.610 & 0.046 & 0.439 & 0.050 & 0.250 & 0.411 \\
 11.500 & 0.697 & 0.076 & 1.100 & 0.040 & 0.369 & 0.043 & 0.260 & 0.358 \\
 12.000 & 0.727 & 0.060 & 0.773 & 0.035 & 0.312 & 0.037 & 0.253 & 0.308 \\
 12.500 & 0.758 & 0.047 & 0.556 & 0.030 & 0.263 & 0.032 & 0.235 & 0.263 \\
 13.000 & 0.788 & 0.037 & 0.404 & 0.025 & 0.220 & 0.027 & 0.210 & 0.221 \\
 13.500 & 0.818 & 0.029 & 0.295 & 0.021 & 0.182 & 0.022 & 0.180 & 0.183 \\
 14.000 & 0.848 & 0.022 & 0.212 & 0.017 & 0.146 & 0.018 & 0.147 & 0.147 \\
 14.500 & 0.879 & 0.016 & 0.149 & 0.013 & 0.113 & 0.014 & 0.115 & 0.113 \\
 15.000 & 0.909 & 0.011 & 0.100 & 0.010 & 0.083 & 0.010 & 0.084 & 0.082 \\
 15.500 & 0.939 & 0.007 & 0.060 & 0.006 & 0.053 & 0.006 & 0.054 & 0.053 \\
 16.000 & 0.970 & 0.003 & 0.027 & 0.003 & 0.026 & 0.003 & 0.026 & 0.026 \\
 16.500 & 1.000 & 0.000 & 0.000 & 0.000 & 0.000 & 0.000 & 0.000 & 0.000\\
\hline
\end{tabular}
\caption{Same as Table~\ref{TBL1} for the SU(3) gauge theory with fundamental fermions.}
\end{table}

\begin{table}[ht!]
\begin{tabular}{ccccccccc}
\hline\multicolumn{9}{c}{SU(4) Fundamental} \\ \hline
$n_f$ & $n_f/\bar{n}_f$ & $A_ 2^*$ & $\gamma _ 2^*$ & $A_ 3^*$ & $\gamma _ 3^*$ & $A_ 4^*$ & $\gamma _ 4^*$ & $\gamma _ {\text{AO}}^*$\\
11.000 & 0.500 & 1.110 & 241.000 & 0.077 & 1.830 & 0.075 & -0.227 & 0.848 \\
 11.500 & 0.523 & 0.463 & 44.000 & 0.068 & 1.400 & 0.067 & -0.053 & 0.775 \\
 12.000 & 0.545 & 0.282 & 17.300 & 0.060 & 1.110 & 0.060 & 0.058 & 0.707 \\
 12.500 & 0.568 & 0.197 & 8.980 & 0.054 & 0.904 & 0.055 & 0.136 & 0.645 \\
 13.000 & 0.591 & 0.147 & 5.380 & 0.048 & 0.755 & 0.050 & 0.192 & 0.587 \\
 13.500 & 0.614 & 0.115 & 3.520 & 0.043 & 0.642 & 0.046 & 0.232 & 0.534 \\
 14.000 & 0.636 & 0.092 & 2.450 & 0.039 & 0.552 & 0.041 & 0.259 & 0.485 \\
 14.500 & 0.659 & 0.075 & 1.770 & 0.035 & 0.480 & 0.038 & 0.275 & 0.439 \\
 15.000 & 0.682 & 0.062 & 1.320 & 0.032 & 0.420 & 0.034 & 0.281 & 0.396 \\
 15.500 & 0.705 & 0.052 & 1.010 & 0.028 & 0.369 & 0.031 & 0.278 & 0.356 \\
 16.000 & 0.727 & 0.043 & 0.778 & 0.025 & 0.325 & 0.027 & 0.269 & 0.318 \\
 16.500 & 0.750 & 0.036 & 0.610 & 0.023 & 0.286 & 0.024 & 0.254 & 0.283 \\
 17.000 & 0.773 & 0.031 & 0.481 & 0.020 & 0.251 & 0.022 & 0.234 & 0.250 \\
 17.500 & 0.795 & 0.026 & 0.380 & 0.018 & 0.218 & 0.019 & 0.211 & 0.218 \\
 18.000 & 0.818 & 0.021 & 0.301 & 0.015 & 0.189 & 0.016 & 0.187 & 0.189 \\
 18.500 & 0.841 & 0.017 & 0.236 & 0.013 & 0.161 & 0.014 & 0.162 & 0.160 \\
 19.000 & 0.864 & 0.014 & 0.183 & 0.011 & 0.134 & 0.012 & 0.136 & 0.134 \\
 19.500 & 0.886 & 0.011 & 0.139 & 0.009 & 0.109 & 0.009 & 0.111 & 0.109 \\
 20.000 & 0.909 & 0.008 & 0.102 & 0.007 & 0.085 & 0.007 & 0.086 & 0.085 \\
 20.500 & 0.932 & 0.006 & 0.071 & 0.005 & 0.063 & 0.005 & 0.063 & 0.062 \\
 21.000 & 0.955 & 0.004 & 0.044 & 0.004 & 0.041 & 0.004 & 0.041 & 0.040 \\
 21.500 & 0.977 & 0.002 & 0.021 & 0.002 & 0.020 & 0.002 & 0.020 & 0.020 \\
 22.000 & 1.000 & 0.000 & 0.000 & 0.000 & 0.000 & 0.000 & 0.000 & 0.000\\
\hline
\end{tabular}
\caption{Same as Table~\ref{TBL1} for the SU(4) gauge theory with fundamental fermions.}
\end{table}

\begin{table}[ht!]
\begin{tabular}{ccccccccc}
\hline\multicolumn{9}{c}{SU(2) Adjoint} \\ \hline
$n_f$ & $n_f/\bar{n}_f$ & $A_ 2^*$ & $\gamma _ 2^*$ & $A_ 3^*$ & $\gamma _ 3^*$ & $A_ 4^*$ & $\gamma _ 4^*$ & $\gamma _ {\text{AO}}^*$\\
1.500 & 0.545 & 0.179 & 5.370 & 0.083 & 1.920 & 0.068 & 1.300 & 1.020 \\
 2.000 & 0.727 & 0.050 & 0.820 & 0.037 & 0.543 & 0.036 & 0.500 & 0.458 \\
 2.500 & 0.909 & 0.011 & 0.139 & 0.010 & 0.127 & 0.010 & 0.127 & 0.122\\
\hline\multicolumn{9}{c}{SU(3) Adjoint} \\ \hline
1.500 & 0.545 & 0.119 & 5.370 & 0.055 & 1.920 & 0.049 & 1.520 & 1.020 \\
 2.000 & 0.727 & 0.033 & 0.820 & 0.024 & 0.543 & 0.025 & 0.523 & 0.458 \\
 2.500 & 0.909 & 0.007 & 0.139 & 0.007 & 0.127 & 0.007 & 0.128 & 0.122\\
\hline\multicolumn{9}{c}{SU(4) Adjoint} \\ \hline
1.500 & 0.545 & 0.089 & 5.370 & 0.041 & 1.920 & 0.038 & 1.620 & 1.020 \\
 2.000 & 0.727 & 0.025 & 0.820 & 0.018 & 0.543 & 0.019 & 0.532 & 0.458 \\
 2.500 & 0.909 & 0.005 & 0.139 & 0.005 & 0.127 & 0.005 & 0.128 & 0.122\\
\hline
\end{tabular}
\caption{Same as Table~\ref{TBL1} for the gauge theories SU(2), SU(3) and SU(4) with adjoint matter.}
\end{table}

\begin{table}[ht!]
\begin{tabular}{ccccccccc}
\hline\multicolumn{9}{c}{SU(3) 2-index Symmetric} \\ \hline
$n_f$ & $n_f/\bar{n}_f$ & $A_ 2^*$ & $\gamma _ 2^*$ & $A_ 3^*$ & $\gamma _ 3^*$ & $A_ 4^*$ & $\gamma _ 4^*$ & $\gamma _ {\text{AO}}^*$\\
1.500 & 0.455 & 0.261 & 23.800 & 0.081 & 5.290 & 0.061 & 3.590 & 1.530 \\
 2.000 & 0.606 & 0.067 & 2.440 & 0.040 & 1.280 & 0.037 & 1.120 & 0.827 \\
 2.500 & 0.758 & 0.025 & 0.639 & 0.020 & 0.474 & 0.020 & 0.466 & 0.407 \\
 3.000 & 0.909 & 0.007 & 0.144 & 0.006 & 0.133 & 0.006 & 0.133 & 0.127\\
\hline\multicolumn{9}{c}{SU(4) 2-index Symmetric} \\ \hline
1.500 & 0.409 & 0.441 & 111.000 & 0.079 & 10.400 & 0.056 & 6.750 & 1.850 \\
 2.000 & 0.545 & 0.077 & 4.820 & 0.039 & 2.080 & 0.035 & 1.790 & 1.060 \\
 2.500 & 0.682 & 0.030 & 1.210 & 0.021 & 0.776 & 0.021 & 0.746 & 0.596 \\
 3.000 & 0.818 & 0.012 & 0.381 & 0.010 & 0.313 & 0.010 & 0.315 & 0.284 \\
 3.500 & 0.955 & 0.002 & 0.064 & 0.002 & 0.062 & 0.002 & 0.062 & 0.061\\
\hline
\end{tabular}
\caption{Same as Table~\ref{TBL1} for the gauge theories SU(3) and SU(4) with 2-index symmetric matter.}
\end{table}

\begin{table}[ht!]
\begin{tabular}{ccccccccc}
\hline\multicolumn{9}{c}{SU(4) 2-index Antisymmetric} \\ \hline
$n_f$ & $n_f/\bar{n}_f$ & $A_ 2^*$ & $\gamma _ 2^*$ & $A_ 3^*$ & $\gamma _ 3^*$ & $A_ 4^*$ & $\gamma _ 4^*$ & $\gamma _ {\text{AO}}^*$\\
5.000 & 0.455 & 4.000 & 4200.000 & 0.089 & 4.150 & 0.130 & -4.940 & 1.190 \\
 5.500 & 0.500 & 0.361 & 38.000 & 0.067 & 2.210 & 0.079 & -0.026 & 0.991 \\
 6.000 & 0.545 & 0.172 & 9.780 & 0.053 & 1.380 & 0.061 & 0.293 & 0.826 \\
 6.500 & 0.591 & 0.105 & 4.170 & 0.043 & 0.952 & 0.049 & 0.402 & 0.686 \\
 7.000 & 0.636 & 0.071 & 2.190 & 0.035 & 0.695 & 0.040 & 0.435 & 0.566 \\
 7.500 & 0.682 & 0.050 & 1.290 & 0.028 & 0.525 & 0.032 & 0.418 & 0.462 \\
 8.000 & 0.727 & 0.036 & 0.802 & 0.023 & 0.402 & 0.025 & 0.368 & 0.372 \\
 8.500 & 0.773 & 0.026 & 0.515 & 0.018 & 0.306 & 0.020 & 0.302 & 0.291 \\
 9.000 & 0.818 & 0.018 & 0.331 & 0.014 & 0.228 & 0.015 & 0.232 & 0.220 \\
 9.500 & 0.864 & 0.012 & 0.206 & 0.010 & 0.161 & 0.010 & 0.164 & 0.156 \\
 10.000 & 0.909 & 0.007 & 0.117 & 0.006 & 0.101 & 0.007 & 0.103 & 0.099 \\
 10.500 & 0.955 & 0.003 & 0.051 & 0.003 & 0.048 & 0.003 & 0.048 & 0.047 \\
 11.000 & 1.000 & 0.000 & 0.000 & 0.000 & 0.000 & 0.000 & 0.000 & 0.000\\
\hline
\end{tabular}
\caption{Same as Table~\ref{TBL1} for the SU(4) gauge theory with 2-index antisymmetric matter.}
\end{table}

\end{document}